# Misfit layered superconductor $(PbSe)_{1.14}(NbSe_2)_3$ with possible layer-selective FFLO state

Yuki M. Itahashi[1,2], Yamato Nohara[2], Michiya Chazono[3], Hideki Matsuoka[1,4], Koichiro Arioka[5], Tetsuya Nomoto[1,6], Yoshimitsu Kohama[6], Youichi Yanase[3], Yoshihiro Iwasa[1,2]*, Kaya Kobayashi[5,7]*

[1] *RIKEN Center for Emergent Matter Science (CEMS), Wako 351-0198, Japan*

[2] *Quantum-Phase Electronics Center (QPEC) and Department of Applied Physics, The University of Tokyo, Tokyo 113-8656, Japan*

[3] *Department of Physics, Graduate School of Science, Kyoto University, Kyoto 606-8502, Japan*

[4] *Institute of Industrial Science, The University of Tokyo, Tokyo 153-8505, Japan*

[5] *Research Institute for Interdisciplinary Science, Okayama University, Okayama 700-8530, Japan*

[6] *The Institute for Solid State Physics, The University of Tokyo, Kashiwa 277-8581, Japan*

[7] *Research Institute for Electronic Science, Hokkaido University, Sapporo 001-0020, Japan*

*Corresponding author: iwasay@riken.jp, kayakobayashi@es.hokudai.ac.jp

**Two-dimensional (2D) superconductors are known for their novel emergent phenomena, however, lack of experimental probes beyond resistivity has hindered further exploration of diverse superconducting states. Bulk 2D superconductors, with superconducting layers separated by non-superconducting layers, offer a unique opportunity to break this limit. Here, we synthesized a single crystal of misfit layered compound $(PbSe)_{1.14}(NbSe_2)_3$, composed of alternately stacked tri-layer $NbSe_2$ and non-superconducting block layers with incompatible unit cells. Due to its unique structure, 2D Ising superconductivity is**

**maintained even in a bulk form. Resistivity and tunnel diode oscillator measurements reveal two distinct superconducting phases in magnetic field vs. temperature phase diagram. Combined with the theoretical analysis, the high-magnetic-field phase is identified as a layer-selective Fulde-Ferrell-Larkin-Ovchinnikov (FFLO) phase, where Ising and finite-$q$ superconductivity are mixed due to the tri-layer structure. Bulk 2D superconductors with misfit structure offer a novel opportunity for understanding of 2D superconductivity through bulk measurements and interlayer engineering.**

Van der Waals (vdW) layered superconductors have garnered significant attention due to their intriguing properties, such as high-temperature and unconventional superconductivity [1,2] accessibility to the two-dimensional (2D) limit [3–5], and the flexibility of stacking different layers[6]. One key advantage for achieving unconventional superconductivity in these materials is the suppression of the orbital pair-breaking effect under an in-plane magnetic field when reaching the 2D limit. For example, Ising superconductivity has been observed in vdW 2D superconductors, where the upper critical field ($B_{c2}$) for the in-plane field, which is often limited by the Pauli paramagnetic effect, is significantly enhanced by broken inversion symmetry and strong spin-orbit interactions [3,7,8]. More recently, the emergence of Fulde–Ferrell–Larkin–Ovchinnikov (FFLO) phase [9–11], characterized by spatially modulated order parameters, has been experimentally reported [12–14].

The FFLO state has been experimentally observed in many kinds of superconductors [12–24]. Among them, transition metal dichalcogenide (TMD) based superconductors offer a new material platform for investigating the interplay between Ising superconductivity and the FFLO phase. However, in monolayer or ion-gated TMD systems, the FFLO state has not been observed, likely due to insufficient system cleanness. Spatially modulated phases are generally sensitive to randomness. In bulk TMDs, 2H-NbS$_2$ [12] and (Ba$_3$NbS$_5$)$_2$(NbS$_2$)$_9$ [13,25] have been shown to exhibit FFLO states. The latter consists of $Ba_3NbS_5$ block layers and $NbS_2$ superconducting layers. However, Isingness is lost in both systems due to their strong interlayer interactions. Recently, angle-resolved transport measurements have revealed an unconventional FFLO phase in multilayer $NbSe_2$ flakes covered by hexagonal boron nitride [14]. This phase, termed orbital FFLO, arises from the coexistence of system cleanness and Ising superconductivity. However, the micrometer-sized flakes complicate the identification of the phase transitions within the zero-resistance state.

To overcome such a restriction, a misfit layered superconductor $[(MX)_{1+\delta}]_m(TX_2)_n$, where M is metal, X is chalcogen, and T is transition metal atoms, offers ideal material platform. In this material, incommensurate superconducting and non-superconducting block layers are alternately stacked [26–30], keeping the 2D nature even in a bulk form. Especially, Ising superconductivity is maintained in bulk, being called bulk Ising superconductivity [26]. Furthermore, the 2D superconductivity realized in a bulk form enables us to perform additional experiments beyond conventional resistivity measurement, which is crucial for the characterization of unconventional superconducting phases.

In this study, we synthesized single crystals of the misfit layered superconductor $(PbSe)_{1.14}(NbSe_2)_3$, where tri-layer $NbSe_2$ and non-superconducting PbSe with incompatible unit cells are alternately stacked, and explored the superconducting phase diagram. A detailed examination of temperature variations of magnetic-field-dependent resistance reveals an anomalous upturn in the in-plane upper critical field ($B_{c2}$). Additionally, we observed an anomalous component in the angle dependence of the critical field at low temperatures. These results suggest the presence of an unconventional superconducting state in the high magnetic field and low-temperature regime. Tunnel diode oscillator (TDO) measurements further revealed an anomaly that implies a phase boundary between the unconventional and conventional $q = 0$ Ising superconductivity. Based on the theoretical predictions, we identify the unconventional phase as a layer-selective FFLO phase [31], where tri-layer NbSe$_2$ units in the misfit compound triggers mixture of finite-$q$ and spatially uniform superconducting states. Since the upturn in the in-plane critical field is absent in the exfoliated tri-layer nor bulk NbSe$_2$, the cleanness and separation of NbSe$_2$ layers by PbSe block layers play a crucial role in stabilizing this layer-selective FFLO phase.

**Basic properties of (PbSe)$_{1.14}$(NbSe$_2$)$_3$**

Figure 1a shows a crystal structure of (PbSe)$_{1.14}$(NbSe$_2$)$_3$ with the 2H$_a$ stacking of NbSe$_2$ tri-layer, where one Nb atom locates directly on another Nb atom. Each tri−layer NbSe$_2$ stacks mostly in the same direction, which is also experimentally confirmed by the scanning transmission electron microscopy (STEM) measurement (Fig.1b and Supplementary Information section I). The lattice constants of NbSe$_2$ and PbSe layers are nearly commensurate along $b$ axis (armchair direction) and incommensurate along $a$ axis (zigzag direction) [28,30]. It exhibits a superconducting transition at $T_c$ = 4.8 K in a powder sample [28].

We first grew single crystals of (PbSe)$_{1.14}$(NbSe$_2$)$_3$ by chemical vapor transport (see Method). For the transport measurement, we exfoliated the single crystal (PbSe)$_{1.14}$(NbSe$_2$)$_3$ in thin flakes and fabricated micro-size devices (Fig. 1c) with typical thickness of approximately 100 nm. Fig. 1d depicts the temperature dependence of the longitudinal resistivity $\rho_{xx}$ in sample 1. A value of residual resistivity ratio is approximately 12, which is largely enhanced compared to the one in a pelletized polycrystalline sample [28]. Figure 1e shows a magnification of temperature dependent $\rho_{xx}$ around the superconducting transition. The superconducting transition temperature $T_c$, which is defined as the midpoint of the resistive transition, is 5.1 K, while $T_c$ estimated by the bulk magnetic susceptibility measurement is 4.8 K (see Supplemental Information Section II). These values are comparable to that in a powder sample [28]. This indicates that bulk single crystals of (PbSe)$_{1.14}$(NbSe$_2$)$_3$ are successfully synthesized. We also estimated the in-plane mean free path ($l$ = 76.2 nm) and Pippard coherence length ($\xi_P$ = 15.1 nm) and resultant $\xi_P/l$ = 0.198 (see Supplemental Information Section III). This indicates that the present system is relatively clean, having a chance to host spatially modulated superconducting phases.

**Upper critical fields revealed by resistivity measurement**

We first focus on the upper critical fields obtained by the resistivity measurement in $(PbSe)_{1.14}(NbSe_2)_3$. Figures 2a and b show the temperature dependence of longitudinal resistivity $\rho_{xx}$ under several out-of-plane and in-plane magnetic fields, respectively. The superconductivity is robust against the in-plane magnetic field, indicating 2D superconductivity realized in the bulk form. Figures 2c and d indicate the superconducting phase diagram under out-of-plane and in-plane pulsed magnetic field, respectively. Pink and light green circles indicate the critical magnetic field $B_{c2}$, which is defined by the midpoint of the resistive transition. Under out-of-plane magnetic field, $B_{c2}$ is proportional to $T_c-T$. This indicates the orbital pair breaking effect destroys superconductivity. On the other hand, under in-plane magnetic field, we found that $B_{c2}$ is proportional to $(T_c-T)^{1/2}$ around $T_c$, as fitted by brown dashed line. This also indicates the 2D nature of superconductivity. We also note that the in-plane critical field reaches around 40 T at low temperature, which exceeds the Pauli paramagnetic limit $B_P$ = 1.86$T_c$ = 9.5 T. This indicates the realization of the Ising superconductivity, similarly to the case of exfoliated tri-layer NbSe$_2$ [3]. More interestingly, we observed a kink structure in $B_{c2}$ around $T$ = 3 K as shown by black arrow in Fig. 2d. Notably, the observed anomaly in $B_{c2}$ is not seen in either exfoliated tri-layer NbSe$_2$ or bulk NbSe$_2$ [3,32,33]. (see Supplementary Information Section III). Transport data indicates that the present $(PbSe)_{1.14}(NbSe_2)_3$ is cleaner than exfoliated tri-layer NbSe$_2$ and as clean as bulk NbSe$_2$ as shown in Supplementary Table 1. This leads us to conclude that the cleanness of the tri-layer

NbSe2 protected by the PbSe block layers is the key factor for the kink behavior in the $B_{c2}$-$T$ phase diagram.

Next, we discuss the angle dependence of the critical field. In 2D superconductors, the angle dependence of critical field $B_{c2}(\theta)$ obeys following 2D Tinkham formula [34]

$$\left(\frac{B_{c2}(\theta)\sin\theta}{B_{c2}^{ab}}\right)^2 + \left|\frac{B_{c2}(\theta)\cos\theta}{B_{c2}^{c}}\right| = 1, \tag{1}$$

where $B_{c2}^{ab}$ and $B_{c2}^{c}$ are the upper critical field for in-plane and out-of-plane direction, respectively. According to the formula, if we plot $|B_{c2}(\theta)\cos\theta|$ vs $(B_{c2}(\theta)\sin\theta)^2$, they scales linearly with $x$-intercept ($\theta = 90°$) and $y$-intercept ($\theta = 0°$) being $(B_{c2}^{ab})^2$ and $B_{c2}^{c}$, respectively. In Fig. 2e, we plotted $\left|\frac{B_{c2}(\theta)\cos\theta}{B_{c2}(\theta=0°)}\right|$ vs $\left(\frac{B_{c2}(\theta)\sin\theta}{B_{c2}(\theta=90°)}\right)^2$, above (red, $T$ = 4.2 K) and below (blue, $T$ = 0.7 K) the temperature at the kink position, respectively. Here, $B_{c2}(\theta = 90°)$ and $B_{c2}(\theta = 0°)$ are obtained by the linear extrapolation of $|B_{c2}(\theta)\cos\theta|$ vs $(B_{c2}(\theta)\sin\theta)^2$ curves around $x$-intercept ($\theta = 90°$) and $y$-intercept ($\theta = 0°$), respectively. At $T$ = 4.2 K, $\left|\frac{B_{c2}(\theta)\cos\theta}{B_{c2}(\theta=0°)}\right|$ linearly depends on $\left(\frac{B_{c2}(\theta)\sin\theta}{B_{c2}(\theta=90°)}\right)^2$ in all $\theta$ region (black solid line), which indicates the 2D Tinkham formula holds good over the entire region. On the other hand, at $T$ = 0.7 K, $\left|\frac{B_{c2}(\theta)\cos\theta}{B_{c2}(\theta=0°)}\right|$ fails to be linear as a function of $\left(\frac{B_{c2}(\theta)\sin\theta}{B_{c2}(\theta=90°)}\right)^2$. Instead, we need two linear functions to reproduce the data as shown by black dashed and light blue solid lines. This implies that unconventional superconducting phase appears around the in-plane magnetic field direction, making it impossible for $B_{c2}(\theta)$ to be fitted by a single Tinkham formula in entire $\theta$ region (Fig. 2e). Figure 2f show $B_{c2}(\theta)$ at $T$ = 0.7 K. Just around $\theta = 90°$, the angle dependence becomes steeper, resulting in the fitting with two curves: light blue solid line for $|\theta - 90°| < 2°$ with $B_{c2}^{ab} = B_{c2}(\theta = 90°) = 38.0$ T and $B_{c2}^{c} = 1.00$ T, and black dashed line for $|\theta - 90°| > 2°$ with $B_{c2}^{ab} = 29.1$ T and $B_{c2}^{c} = B_{c2}(\theta = 0°) = 1.55$ T. Between black dashed and light blue solid lines, the emergent superconducting phase exists (light blue region). Notably, $B_{c2}^{ab} = 29.1$ T for $|\theta - 90º| > 2º$ coincides with the value of $B_{c2}$ in brown dashed line at $T = 0.7$ K (Fig. 2d), which indicates that black dashed line in Fig. 2f describes the angle dependence of $B_{c2}$ without the unconventional phase.

These anomalies in temperature and angle dependencies of upper critical field are manifestation of unconventional superconductivity at low temperature and high magnetic field region. Generally speaking, the spatially uniform $q = 0$ state is unstable in the magnetic field since Zeeman splitting and orbital effect prevent a formation of singlet Cooper pairs. However, by acquiring spatially modulated order parameter, superconductivity can survive even under high magnetic field. Examples include FFLO phase [9,10], helical/complex stripe phase in noncentrosymmetric/locally noncentrosymmetric superconductors [35–39], pair density wave (PDW) phase in nonsymmorphic superconductors [37,40–43], and orbital FFLO phase caused by in-plane vortices [14,37,44]. These spatially modulated superconducting phases are stabilized by the applied magnetic field and the phase boundary inside the zero-resistance state can be detected by thermodynamic or magnetic susceptibility measurements [12,16,18,22–24].

**Superconducting phase diagram drawn by TDO measurement**

To clarify the occurrence of a phase transition between the conventional and spatially modulated superconducting states, we performed tunnel diode oscillator (TDO) experiment [45] in bulk $(PbSe)_{1.14}(NbSe_2)_3$ under in-plane magnetic field (Fig. 3). TDO measurement is a highly sensitive probe of the emergence or change in the superconducting state, enabling us to find the phase boundary inside the zero-resistance state [18–21,24]. Figures 3a and b show the TDO frequency difference $\Delta F$ as a function of in-plane magnetic field at $T = 2.00$ K and various temperatures, respectively. Above $T_c$ at $T = 6.08$ K (black curve in Fig. 3b), $\Delta F$ gradually decreases with increasing $B$ due to the background signal of our setup. On the other hand, below $T_c$ (curves except for black one in Figs. 3a and b), $\Delta F$ shows sharp decrease with kinks with increasing $B$ due to suppression of superconductivity. Figures 3c and d show the first derivative of $\Delta F$ as a function of $B$ ($dF/dB$) at $T = 2.00$ K and various temperatures, respectively. Below $T_c$, $dF/dB$ curves show large dip structures shown by gray squares ($B_1$) despite the zero-resistance at $B_1$ according to Fig. 2d. We assume that $B_1$ indicates the melting of Josephson vortices [46] induced in the PbSe layers. In insulating layers sandwiched by superconducting layers, Josephson vortices are induced under in-plane magnetic field [47], whose circular supercurrent flows through the superconducting layers horizontally and the insulating layer vertically. Josephson vortices form a vortex lattice in insulating layers at low magnetic field, and with increasing magnetic field, the vortex lattice melts [17,48] or changes into uniform magnetic field in PbSe layers. This melting can be detected by TDO as a change in the effective penetration depth [46], but is not captured by the in-plane resistance measurement [17]. When the in-plane vortices are driven by current, it moves toward out-of-plane direction. However, the superconducting layers prevent the out-of-plane motion of vortices and zero-resistance state is maintained although the Josephson vortex lattice is melted. Thus, at $B_1$, the zero-resistance is realized, while the anomaly appears in the TDO measurement.

Here, we discuss another anomaly in the superconducting state. In Figs. 3e and f, we plot the second derivative of $\Delta F$ as a function of $B$ ($d^2F/dB^2$) at $T = 2.00$ K and various temperatures, respectively. At low temperatures, we can see two peak structures at high and low fields, which are denoted by light green diamonds and pink circles, respectively. With increasing temperature, the lower-field peaks (light green diamonds) merge to the higher-field peaks (pink circles). These can be also seen as one or two step structures in $dF/dB$. Above the magnetic field at the higher-field peak position (pink circles), $dF/dB$ shows constant values and the values of $\Delta F$ are comparable to that at $T = 6.08$ K (black) under high fields, i.e., in the normal state. Thus, that magnetic field should correspond to the upper critical field ($B_{c2}$). The magnetic field $B_2$ at the lower-field peak indicated by light green diamonds, should be a phase boundary between conventional and unconventional superconducting state. We note that the value of $B_2$ (~27 T) is consistent with $B_{c2}^{ab} = 29.1$ T estimated from the fitting (dashed line) in the range of $|\theta - 90°| > 2°$ at $T = 0.7$ K (Fig. 2f), which further ascertain that $B_2$ separates the two superconducting phases.

Figure 4 summarizes the superconducting phase diagram under in-plane magnetic field obtained by TDO measurement. We can see kink structure in $B_{c2}$ around $T = 3$ K similarly to the resistivity measurement (Fig. 2d). Just below the temperature at the kink position, $B_2$ appears and shows almost constant value toward low temperature, forming the phase boundary between the lower field region below $B_2$ and the higher field region (light blue region). $B_1$ is assigned as the vortex melting line of Josephson vortices in PbSe layers.

**Discussion**

We now consider the possible order parameters in the present $(PbSe)_{1.14}(NbSe_2)_3$ under in-plane magnetic field. Since each NbSe$_2$ tri-layer has a so-called Zeeman type band splitting even under zero magnetic field [32], the low-field region is likely assigned as the Ising superconductivity. Thus, the conventional FFLO picture is no longer valid for the high field phase. Also, the helical phase in noncentrosymmetric superconductors is less possible because Zeeman type spin splitting is not coupled to the in-plane magnetic field. Thus, we must consider other exotic superconducting phases. As we pointed out from the absence of anomaly in $B_{c2}$ in bulk NbSe$_2$ (see Supplementary Information Section IV), separation of tri-layer NbSe$_2$ segments by block layers is crucial for the emergence of the high-field and low-temperature phase.

We then performed a theoretical calculation of superconducting instability in tri-layer NbSe$_2$ (see Supplementary Information section V) [31] in a model reproducing the band structure and Ising spin-orbit coupling in NbSe$_2$ [49]. Figure 5 shows the *B-T* phase diagram and the order parameters of each region. The filled circles and open squares show $B_{c2}$s of the finite-*q* and *q* = 0 superconducting states, respectively. In contrast to the experimentally obtained $B_{c2}$ (Fig. 4), theoretically calculated $B_{c2}$s diverge toward low temperature. This behavior generally appears in clean 2D superconductivity and is absent in the dirty case [50]. In the experiment, the effect of impurities and defects cannot be neglected, resulting in the saturation of $B_{c2}$ toward low temperature.

The calculation confirmed that the spatially uniform Ising superconductivity is stabilized in the low magnetic field region (light blue region). The order parameter for each layer is not necessarily equal as shown in the left of Fig. 5. On the other hand, in the narrow region between filled dots and open squares, the finite-*q* orbital FFLO phase is realized (pink region). Here, the top and bottom NbSe$_2$ monolayers have Fulde–Ferrell (FF)-like order parameters, and a proximately induced Larkin–Ovchinnikov (LO)-like order parameter is induced in the middle NbSe$_2$ monolayer, as shown by the dashed arrows (right of Fig. 5). Between the Ising superconducting state and the orbital FFLO phase, the layer-selective FFLO phase, where *q* = 0 and finite-*q* order parameters are mixed, is expected (violet region). In addition to the FF- and LO-like order parameters, the middle NbSe$_2$ layer prefers a uniform order parameter, which is proximately induced in the top and bottom NbSe$_2$ layers, as indicated by the dashed arrows of top of Fig. 5. Thus, uniform and spatially modulated order parameters are mixed. Furthermore, the calculation of the real-space gap structure reveals the existence of a phase boundary between the Ising superconductivity and the layer-selective FFLO phase starting from the kink position (Black dashed line). However, the calculation of temperature dependence of the phase boundary still remains a challenge [31]. Based on this theoretical consideration, we attributed the experimentally observed $B_2$ in Fig. 4 to the phase boundary between the Ising superconductivity and the layer-selective FFLO phase (black dashed line), where the finite-$q$ order parameter appears. Since the black circles (phase boundary between the layer-selective FFLO phase and the orbital FFLO phase) and white squares (phase boundary between the orbital FFLO phase and the normal state) are too close in the theoretical calculation to distinguish, these two-phase boundaries should be observed as a single boundary at $B_{c2}$ in Fig. 4. We note that the above discussion is valid because the sets of tri-layer NbSe$_2$ layers are well separated by PbSe layers; a configuration made possible by the unique structure of the present bulk 2D system.

**Conclusion and perspective**

In single crystals of the misfit layered superconductor (PbSe)$_{1.14}$(NbSe$_2$)$_3$, we found the phase transition inside the zero-resistance state by the resistivity and TDO measurements, and the theoretical consideration implies that the high field phase is a layer-selective FFLO phase. Verification of the layer dependent superconducting order parameters by a layer-selective bulk measurement probes, such as nuclear magnetic resonance, is now a future study toward the comprehensive understanding of the tri-layer superconductor.

This class of materials is a unique system with well-defined numbers of superconducting layers incorporated in a bulk material, which will enrich our understanding of 2D superconductivity from a different angle, employing inter-layer engineering and bulk measurements.

## Methods

### Sample preparation

Powder of Pb (3N Kojundo), Nb (3N Furuchi), and Se (3N Kojundo) are weighed with the molar ratio of Pb: Nb: Se = 1: 2: 5. Pelletized mixture was sealed in an evacuated quartz tube ($10^{-5}$ Torr) and heated at 900 °C for a day. 300 mg of obtained charge and 20 mg of $TeCl_4$ was sealed in an evacuated quartz tube ($10^{-2}$ Torr) and heated with a gradient of 50 °C when the high temperature is set to 800 °C. After a week, we obtained single crystals of $(PbSe)_{1.14}(NbSe_2)_3$. The stacking sequence was characterized by the X-ray diffraction pattern for both sides to avoid the contamination of alternating stackings.

For the resistivity measurement, the obtained $(PbSe)_{1.14}(NbSe_2)_3$ single crystals were exfoliated into thin flakes using the Scotch-tape method, and the flakes were transferred onto a $Si/SiO_2$ substrate. The thickness of the exfoliated flakes was measured using atomic force microscopy. A Hall bar configuration was fabricated on the flakes with Au (150 nm)/Ti (9 nm) electrodes. The pattern was fabricated using electron beam lithography, and the electrodes were deposited using an evaporator.

### Transport measurement

In Quantum Design Physical Property Measurement System, the resistance was measured using AC lock-in amplifiers (Stanford Research Systems Model SR830 DSP) with a frequency of 13 Hz. In high magnetic field, the magnetoresistance was measured using a numerical lock-in technique operating at 50 kHz. The high magnetic fields were generated by a pulsed magnet which provides a peak field of 48 T with a total field duration of 36 ms. The angle between $(PbSe)_{1.14}(NbSe_2)_3$ single crystalline sample and magnetic field was tuned by a plastic single axis rotator. The temperature was controlled between 0.6 K and 4 K with a homemade plastic $^3$He refrigerator.

**TDO measurement**

The angle dependent tunnel diode oscillator (TDO) measurements were conducted in pulsed magnetic fields by attaching a (PbSe)$_{1.14}$(NbSe$_2$)$_3$ single crystalline sample on a double counter would pick-up coil. The coil was a part of the self-resonance circuit made of the tunnel diode and was resonated at ~ 80 MHz. The change in the superconducting properties was observed as a shift of the resonance frequency [45].

**Data availability:** All data needed to evaluate the conclusions of this study are available as Supplementary Information.

**Acknowledgments**

We thank Y. Xie, N. Nagaosa, S. Uji, S. Sugiura, and T. Shibauchi for fruitful discussions. Y.M.I. was supported by JSPS KAKENHI Grant No. JP24K17008. M.C. is supported by JSPS KAKENHI Grant No. JP23KJ1219. M.H. is supported by JSPS KAKENHI Grant No. 21K13888. T.N. is supported by JSPS KAKENHI Grant No. 23K13054. Y.K. is supported by JSPS KAKENHI Grant Nos. 23K17666, 23KK0052, and 22H00104. Y.Y. is supported by JSPS KAKENHI Grant Nos. JP22H01181, JP22H04933, JP23K22452, JP23K17353, JP24H00007, and JP24K21530. Y.I. is supported by JSPS KAKENHI Grant No. JP19H05602. K.K. is supported by JSPS KAKENHI Grant No. 23H04866.

**Author contributions**

Y.M.I., Y.I., and K.K. conceived the research project. K.A. and K.K. synthesized the bulk material. Y.N. fabricated the microdevices, performed the transport measurement in PPMS, and analyzed the data. Y.N., H.M., and Y.M.I. performed high-magnetic-field measurement of transport properties and TDO with the experimental help of T.N. and Y.K. Y.M.I., M.C. Y.K., Y.Y, Y.I., and K.K. wrote the manuscript. All authors have led the physical discussions.

**Competing financial interests**

The authors declare no competing interests.

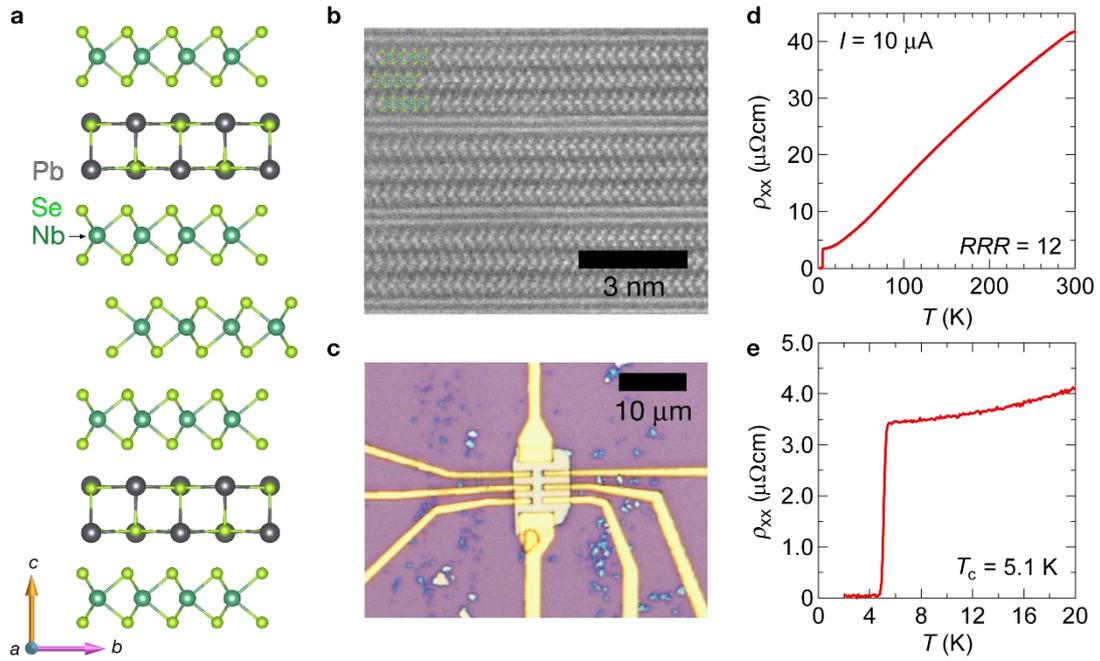

**Fig. 1 | Crystal structure and superconducting properties of (PbSe)$_{1.14}$(NbSe$_2$)$_3$ microdevice.**

**a**, Schematic crystal structure of (PbSe)$_{1.14}$(NbSe$_2$)$_3$. PbSe layers are intercalated in tri-layer NbSe$_2$ with 2H$_a$ stacking. **b**, A cross-sectional scanning transmission electron microscope (STEM) image of (PbSe)$_{1.14}$(NbSe$_2$)$_3$. Schematics of the cross-sectional image of tri-layer 2H$_a$-NbSe$_2$ is also depicted. **c**, Optical image of (PbSe)$_{1.14}$(NbSe$_2$)$_3$ microdevice (sample 1). **d**, Temperature dependence of the resistivity $\rho_{xx}$ in sample 1 at $I = 10$ μA and $B = 0$ T. The residual resistivity ratio RRR = $\rho_{xx}(T = 300$ K$)/\rho_{xx}(T = 6$ K$)$ was 12. **e**, Magnification of temperature dependence of $\rho_{xx}$ around the superconducting transition ($T_c = 5.1$ K).

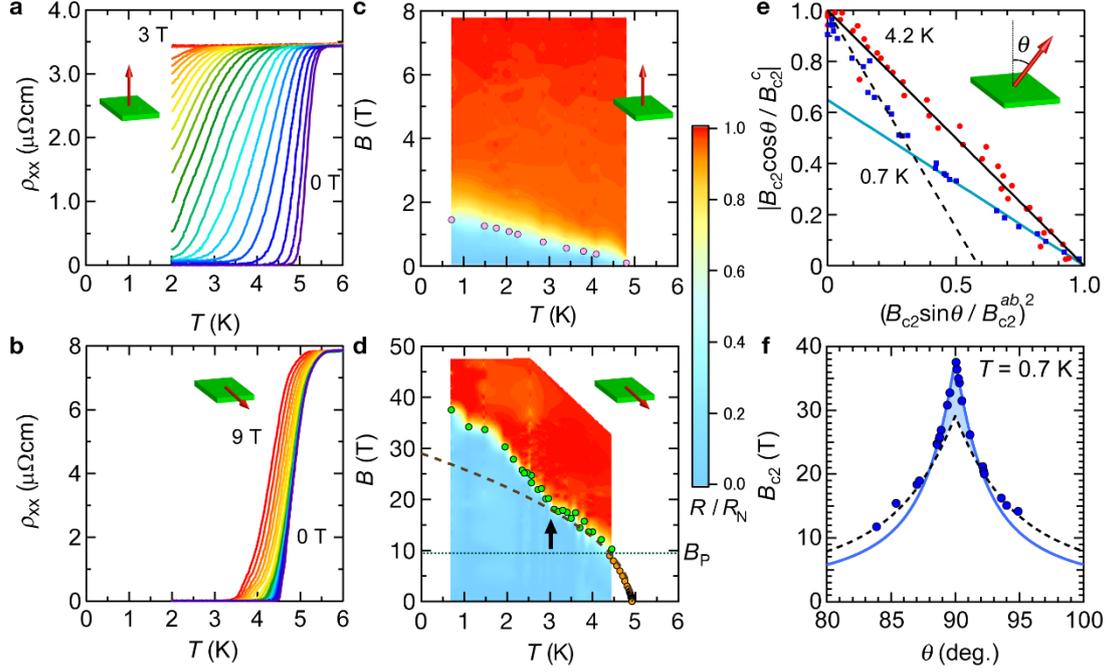

**Fig.2 | Temperature and angle dependence of critical magnetic field. a**, longitudinal resistivity ($\rho_{xx}$) as a function of temperature $T$ in sample 1 under various out-of-plane magnetic fields at $I$ = 50 μA. The magnetic field $B$ was varied in 0.1-T steps from 0 to 2 T and in 0.5-T steps from 2.5 to 3 T. **b**, $\rho_{xx}$ as a function of $T$ in sample 2 under various in-plane magnetic fields at $I$ = 50 μA. $B$ was varied in 0.2-T steps from 0 to 2 T, in 0.5-T steps from 2.5 to 3 T, and 1-T steps from 4 to 8 T. **c, d**, Phase diagram in the $B$-$T$ plane with $B$ along out-of-plane (C) and in-plane (D) directions drawn by the magnitude of $R/R_N$ in sample 2. Pink and light green circles represent the critical field ($B_{c2}$) defined as $R/R_N$ = 0.5. Black arrow in Fig. d indicates the position of upturn, which indicates existence of unconventional superconducting state. Color mapping displays the magnitude of $R/R_N$ obtained from pulsed magnetic field measurement. Brown dashed line is the fitting curve by 2D Tinkham model $B_{c2} = A(1-T/T_c)^{1/2}$, where $A$ = 29.0 T. **e**, $\left|\frac{B_{c2}(\theta)\cos\theta}{B_{c2}(\theta=0°)}\right|$ vs $\left(\frac{B_{c2}(\theta)\sin\theta}{B_{c2}(\theta=90°)}\right)^2$ at $T$ = 4.2 K (sample 1) and $T$ = 0.7 K (sample 2), where $B_{c2}(\theta)$ is angle dependent upper critical field and $\theta$ is angle. Inset shows the definition of $\theta$. $B_{c2}(\theta = 90°)$ and $B_{c2}(\theta = 0°)$ are obtained by the linear extrapolation of $|B_{c2}(\theta)\cos\theta|$ vs $(B_{c2}(\theta)\sin\theta)^2$ curves around x-intercept ($\theta$ = 90°) and y-intercept ($\theta$ = 0°),

respectively. Here, $B_{c2}(\theta = 90°)$ and $B_{c2}(\theta = 0°)$ are 5.03 T and 0.155 T (38.0 T and 1.55 T) at $T$ = 4.2 K ($T$ = 0.7 K), respectively. Black solid line is linear fitting at $T$ = 4.2 K. Black dashed and light blue solid lines are linear fitting at $T$ = 0.7 K around $\theta$ = 90° and $\theta$ = 0°, respectively. **f**, $\theta$ dependence of $B_{c2}$ at $T$ = 0.7 K in sample 2. Black dashed (blue) line indicates the fitting curve by eq (1) in the region of $|\theta - 90°| > 2°$ ($|\theta - 90°| < 2°$). Light blue region depicts the finite-$q$ layer-selective FFLO phase.

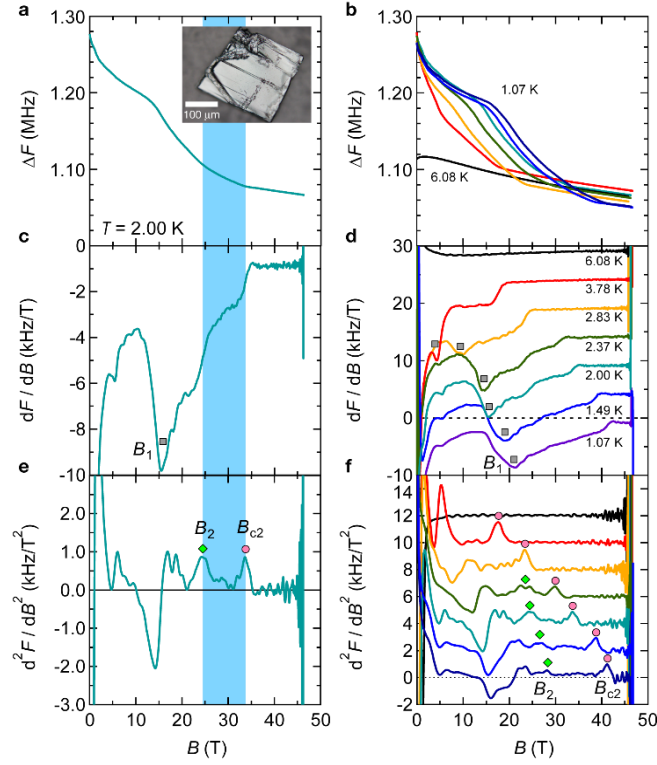

**Fig. 3 | TDO measurement in bulk (PbSe)$_{1.14}$(NbSe$_2$)$_3$ and superconducting phase diagram. a**, TDO frequency change $\Delta F$ as a function of $B$ at $T = 2.00$ K. Inset shows the image of bulk (PbSe)$_{1.14}$(NbSe$_2$)$_3$. **b**, $\Delta F$ as a function of $B$ at $T = 1.07$ K (purple), 1.49 K (blue), 2.00 K (turquoise), 2.37 K (green), 2.83 K (orange), 3.78 K (red), and 6.08 K (black). **c**, First derivative of $\Delta F$ as a function of $B$ (d$F$/d$B$). Gray square indicates the position of the magnetic field at the dip position of d$F$/d$B$, which is vortex melting field $B_1$. **d**, $B$ dependence of d$F$/d$B$ at various temperatures. Each curve is shifted vertically by 5 kHz/T for clarity. Gray squares indicate $B_1$. **e**, Second derivative of $\Delta F$ as a function of $B$ (d$^2F$/d$B^2$). Pink circle and light green diamond indicate the position of the magnetic field at the peak position of d$^2B$/d$F^2$ at higher and lower fields, which is upper critical field $B_{c2}$ and critical field between $q = 0$ Ising superconductivity and the layer-selective FFLO phase $B_2$, respectively. **f**, $B$ dependence of d$^2F$/d$B^2$ at various temperatures. Each curve is shifted vertically by 2 kHz/T$^2$ for clarity. Pink circles and light green diamonds indicate $B_{c2}$ and $B_2$, respectively.

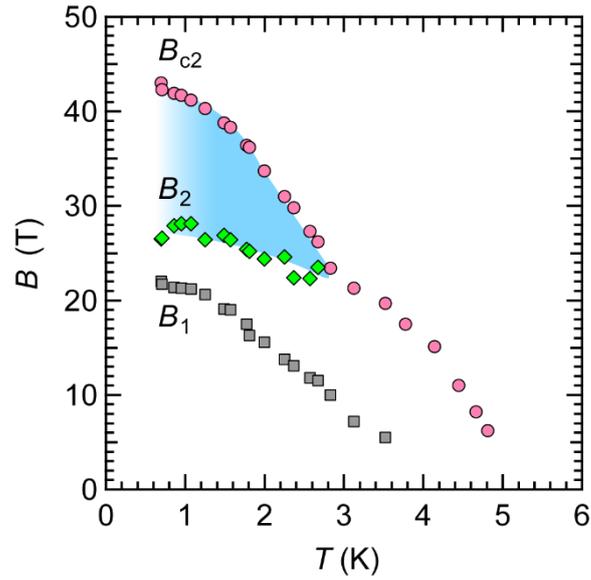

**Fig. 4 | Superconducting phase diagram of PbSe$_{1.14}$(NbSe$_2$)$_3$.** Phase diagram in *B-T* plane for PbSe$_{1.14}$(NbSe$_2$)$_3$ determined from TDO measurement (Fig. 3). Pink circles, light green diamonds, and gray squares indicate upper critical field $B_{c2}$, critical field $B_2$ between $q = 0$ Ising superconducting and layer-selective FFLO phase, and vortex melting field $B_1$, respectively. Light blue region depicts the layer-selective FFLO phase.

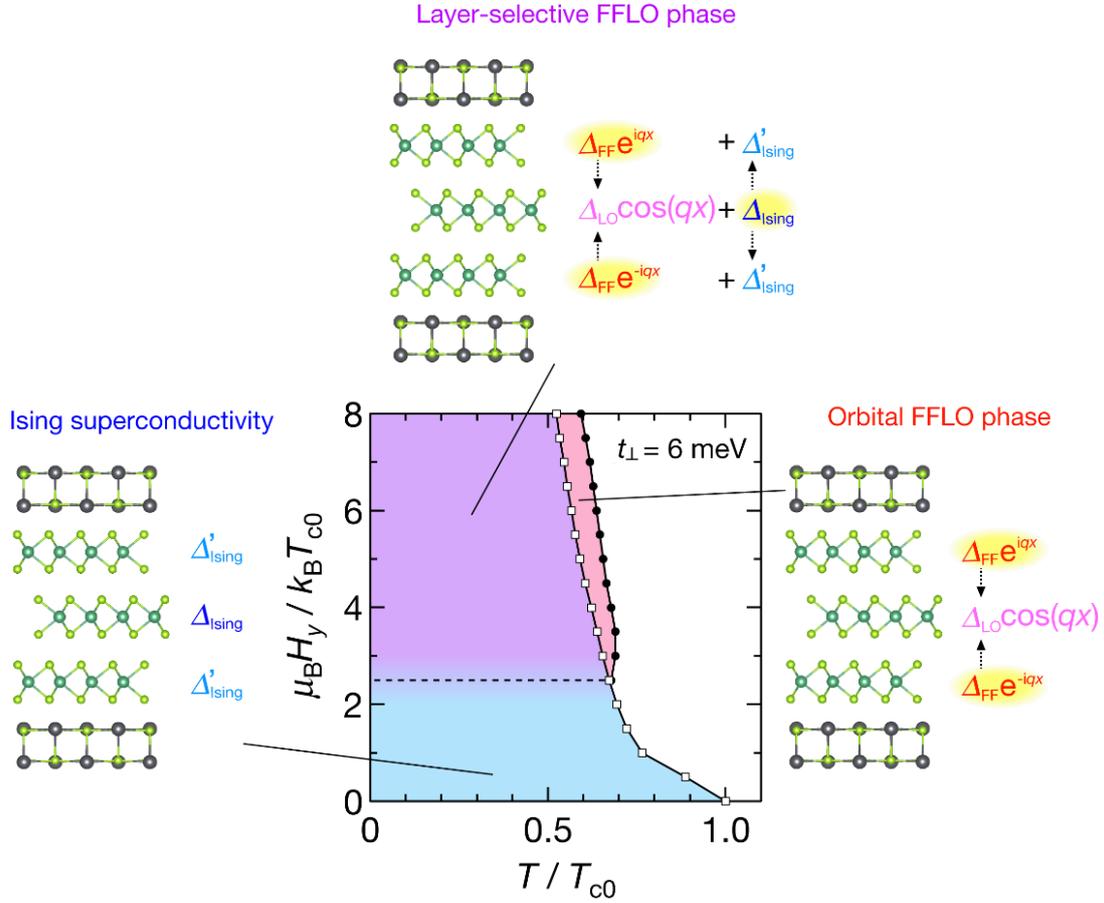

**Fig. 5 | Theoretical calculation of superconducting phase diagram in tri-layer NbSe$_2$.** Black circles and white squares indicate the superconducting transition temperature of finite-$q$ orbital FFLO phase (right) and $q = 0$ Ising superconducting state (left), respectively. These two phases are mixed in a high magnetic field and low temperature region, which is termed layer-selective FFLO phase (top). The dashed line indicates an existing phase boundary between the Ising superconductivity and the layer-selective FFLO phase. The orbital FFLO phase, spatially uniform Ising superconductivity, and the layer-selective FFLO phase are realized in pink, light blue, and violet regions, respectively. The dashed arrows indicate the proximate induction of the order parameter to the adjacent layer.